\documentclass[%
 reprint,
superscriptaddress,
nofootinbib,
 amsmath,amssymb,
 aps,
pra,
]{revtex4-2}

\usepackage{graphicx}
\usepackage{dcolumn}
\usepackage{bm}
\usepackage[hidelinks]{hyperref}
\usepackage{float}
\usepackage{booktabs}
\usepackage{textcase}
\usepackage{makecell} 

\usepackage[locale=US,mode=math]{siunitx}
\DeclareSIUnit\liter{l}
\DeclareSIUnit\bar{bar}

\usepackage[capitalize]{cleveref} 
\usepackage[version=4]{mhchem}
\usepackage{threeparttable} 
\begin{document}

\title{Additively manufactured ceramics for compact quantum technologies}

\author{Marc Christ}
 \email{marc.christ@fbh-berlin.de}
 \affiliation{Ferdinand-Braun-Institut (FBH), Gustav-Kirchhoff-Straße 4, 12489 Berlin, Germany}
  \affiliation{Institut f{\"u}r Physik, Humboldt-Universit\"at zu Berlin, Newtonstraße 15, 12489 Berlin, Germany}
\author{Conrad Zimmermann}
 \affiliation{Ferdinand-Braun-Institut (FBH), Gustav-Kirchhoff-Straße 4, 12489 Berlin, Germany}
\author{Sascha Neinert}
 \affiliation{Ferdinand-Braun-Institut (FBH), Gustav-Kirchhoff-Straße 4, 12489 Berlin, Germany}
  \affiliation{Institut f{\"u}r Physik, Humboldt-Universit\"at zu Berlin, Newtonstraße 15, 12489 Berlin, Germany}
\author{Bastian Leykauf}
  \affiliation{Institut f{\"u}r Physik, Humboldt-Universit\"at zu Berlin, Newtonstraße 15, 12489 Berlin, Germany}
\author{Klaus D\"oringshoff}
 \affiliation{Ferdinand-Braun-Institut (FBH), Gustav-Kirchhoff-Straße 4, 12489 Berlin, Germany}
  \affiliation{Institut f{\"u}r Physik, Humboldt-Universit\"at zu Berlin, Newtonstraße 15, 12489 Berlin, Germany}
\author{Markus Krutzik}
 \affiliation{Ferdinand-Braun-Institut (FBH), Gustav-Kirchhoff-Straße 4, 12489 Berlin, Germany}
  \affiliation{Institut f{\"u}r Physik, Humboldt-Universit\"at zu Berlin, Newtonstraße 15, 12489 Berlin, Germany}

\date{\today}

\begin{abstract}
Quantum technologies are advancing from fundamental research in specialized laboratories to practical applications in the field, driving the demand for robust, scalable, and reproducible system integration techniques. Ceramic components can be pivotal thanks to high stiffness, low thermal expansion, and excellent dimensional stability under thermal stress. We explore lithography-based additive manufacturing of technical ceramics especially for miniaturized physics packages and electro-optical systems. This approach enables functional systems with precisely manufactured, intricate structures and high mechanical stability while minimizing size and weight. It facilitates rapid prototyping, simplifies fabrication and leads to highly integrated, reliable devices. As an electrical insulator with low outgassing and high temperature stability, printed technical ceramics such as \ce{Al2O3}  and \ce{AlN} bridge a technology gap in quantum technology and offer advantages over other printable materials. We demonstrate this potential with CerAMRef, a micro-integrated rubidium D2 line optical frequency reference on a printed \ce{Al2O3} micro-optical bench and housing. The frequency instability of the reference is comparable to laboratory setups while the volume of the integrated spectroscopy setup is only \SI{6}{\milli\liter}. We identify potential for future applications in compact atomic magnetometers, miniaturized optical atom traps, and vacuum system integration. 
\end{abstract}

\maketitle

\section*{Additive manufacturing for compact quantum technologies}
Atomic quantum sensors exploit the quantum properties of atoms to measure physical quantities like gravity, electric and magnetic fields, inertial forces, and time with unparalleled precision and accuracy~\cite{Kitching2011, Bongs2019}. These sensors therefore have significant potential in advancing navigation, timekeeping, and resource exploration. Thanks to their precision, they can be central in fundamental physics research or Earth observation in both terrestrial and extraterrestrial settings~\cite{Becker2018a, Canuel2020, Frye2021, Leveque2023, Abend2023}. 
These varying applications require devices with high level of system integration, suitable for practical field and space deployment. This demands the extensive integration and miniaturization of all subsystems, including the laser system, electronics and the physics package. Additionally, these devices must be robust against environmental influences such as mechanical stress, thermal fluctuations, and radiation exposure to ensure reliability on mobile platforms. Future commercialization and small-series production add further requirements: systems and their components must be reproducible, scalable and cost-effective to produce and easy to assemble.

\begin{figure}[tb]
	  \includegraphics[width=\linewidth]{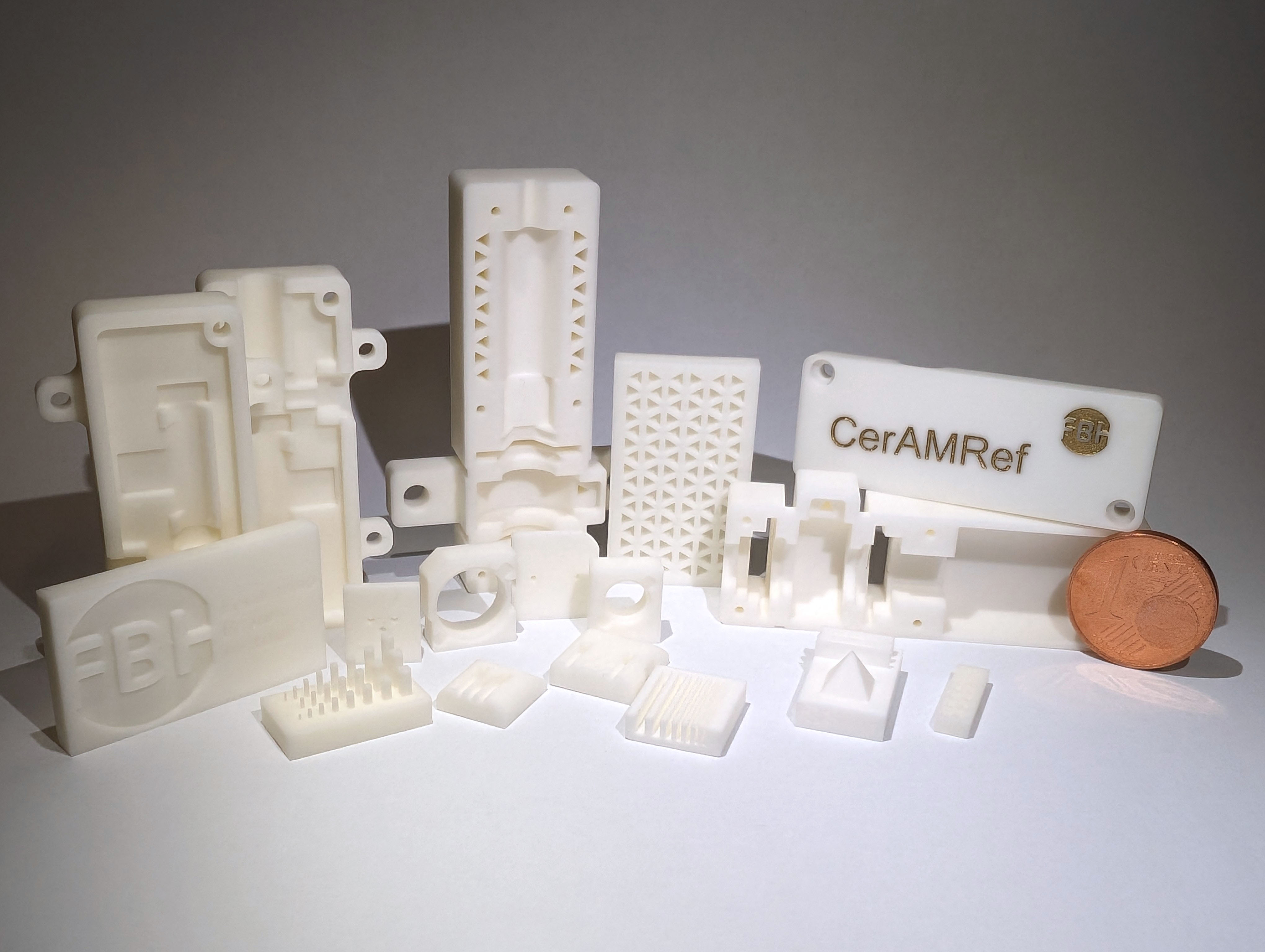}
    \caption{\label{fig:example_components} Selection of additively manufactured \ce{Al2O3} components, including micro-optical benches, housing lids, isogrids, holders for electro-optical components, and test structures for resolution and process parameters.}
\end{figure}

Additive manufacturing (AM) is an innovative approach particularly beneficial for realizing compact physics packages for quantum technologies. A physics package is the core assembly containing the quantum system and is tailored to the specific application. For atomic quantum sensors, it usually incorporates atoms within a vapor cell or high vacuum system and electro-optical systems for state preparation and readout. AM enables a higher level of device integration and functionality, offering increased freedom for functional designs \cite{Madkhaly2021}. Initial applications of AM in quantum technologies have primarily involved 3D printing of metal components for vacuum systems. Notable examples include vacuum components such as resistively heated alkali metal dispenser (Ti-6Al-4V)~\cite{Norrgard2018a}, magnetic coils (Al-Si10-Mg)~\cite{Saint2018} and shields (permalloy-80)~\cite{Vovrosh2017}. Vacuum flanges (Ti-6Al-4V)~\cite{Vovrosh2017} and a compact vacuum chamber (Al-Si10-Mg)~\cite{Cooper2021} with gyroid structures for weight reduction have been realized and demonstrate the suitability of additively manufactured metal components for vacuum systems in quantum technology. Additionally, an optical spectroscopy system for cold atom applications (volume \SI{89}{\milli\liter}, based on \SI{12.7}{\milli\metre} optics) has been realized using a photopolymer as the base material for mounting optical components~\cite{Madkhaly2022}.

This work investigates additively manufactured ceramics for applications in miniaturized quantum technologies, exemplary components are displayed in~\cref{fig:example_components}.

\section*{3D printed ceramics for miniaturized physics packages and electro-optical systems}\label{sec:ceramicAM}

Transferring quantum sensors and electro-optical systems to compact and robust devices for field applications requires suitable materials, miniaturized components and qualified assembly and integration technologies. Important physical and thermal properties of selected materials for compact quantum technologies are summarized in \cref{tab:material_properties}. This includes printable materials along with others commonly used in miniaturized physics packages and electro-optical systems. For mobile applications, selecting lightweight materials with a high Young's modulus \(E\) is critical to guarantee robust setups with minimal deformations under environmental loads. For electro-optical systems, matching the coefficient of thermal expansion (CTE) to other commonly used materials (like BK7 or titanium) is crucial for ensuring stable connections. High thermal conductivity \({\kappa}\) is essential to maintain minimal thermal gradients, thereby reducing non-planar deformations. The assembled setups must also meet common thermal qualification standards for mobile devices \cite{MILSTD883L10108A}, which require a minimum temperature stability from \SI{-55}{\celsius} to \SI{85}{\celsius}. Vacuum compatibility is another crucial criterion for use within hermetically sealed devices or in the vacuum systems of physics packages. 

\begin{table*}[tb]
\caption{\label{tab:material_properties} Relevant physical properties of materials for applications in miniaturized physics packages and electro-optical systems: Young's modulus \(E\), density \(\rho\), coefficient of thermal expansion (CTE), thermal conductance \({\kappa}\) at room temperature. Furthermore, the temperature and vacuum suitability and machinability of the materials in standard machine shops is denoted. The table is divided into two sections: above the line are materials with established printing processes including popular engineering polymers, and below are materials typically used in sub-assemblies such as micro-optical systems. Ceramics are highlighted in bold. Where available, typical properties of printed materials are provided. Note that these properties vary based on test method, manufacturing machine, process parameters, part geometry, and post-processing. They should be regarded as reference values and require experimental validation for specific applications.}
\begin{threeparttable}
\begin{ruledtabular}
\begin{tabular}{@{}l S[table-format=3.2, round-mode=places, round-precision=2] S[table-format=3.1, round-mode=places, round-precision=0] S[table-format=2.2, round-mode=places, round-precision=1,separate-uncertainty=true, multi-part-units=single,table-space-text-pre={$\pm$}]  S[table-format=3.0, round-mode=places, round-precision=1] c c l@{}}
   \multicolumn{1}{@{}l}{Material} &     
    \multicolumn{1}{c}{\thead{\(\rho\) \\ $\left[\si[detect-weight, mode=text]{\gram\per\cubic\centi\metre}\right]$}} &     
    \multicolumn{1}{c}{\thead{\(E\) \\ $\left[\si{\giga\pascal}\right]$}} &  
    \multicolumn{1}{c}{\thead{CTE \\ $\left[10^{-6} \si[detect-weight, mode=text]{\per\kelvin}\right]$}} &    
    \multicolumn{1}{c}{\thead{\(\kappa\) \\ $\left[\si[detect-weight, mode=text]{\watt\per\meter\per\kelvin}\right]$}} & 
    \multicolumn{1}{c}{\thead{Temp. \& vac. \\ suitability\tnote{a}}} &
    \multicolumn{1}{c}{\thead{Machinability\tnote{b}}} &      
    \multicolumn{1}{l@{}}{Ref.} \\
    \hline
    \textbf{\ce{Al2O3}} & 3.985 & 380 & 7.6 & 29 & \checkmark & $-$ & \cite{Lithoz_Mat} \\
    \textbf{\ce{AlN}} & 3.26 & 308 & 3.9 & 160 & \checkmark & $-$ & \cite{AlN, Rauchenecker2022}\\
    Al-Si10-Mg & 2.68 & 70 & 23 & 130 & \checkmark & $++$ & \cite{AlSi10Mg}\\
    Rigid 4000 & 1.26 & 4.1 & 63 &   0.3  & {$\times$}\tnote{c} & $++$  & \cite{Rigid4000, Lee2021}\tnote{d}\\
    High Temp & 1.14 & 2.8 & 75 &   0.3  & {$\times$}\tnote{c} & $++$  & \cite{High_Temp, Lee2021}\tnote{d}\\
    PLA & 1.25 & 2.3 & 68 &  0.18 & {$\times$}\tnote{e} & $++$  & \cite{Spinelli2022, PLA, Madkhaly2022} \\
    SS 316L & 7.98 & 180 & 16 & 15 & \checkmark & $+$ & \cite{316L}\\
    Ti-6Al-4V & 4.47 & 110 & 9 & 6.6 & \checkmark & $\circ$ & \cite{TiAl6V4} \\
    \textbf{\ce{ZrO2}} & 6.088 & 205 & 10 & 2.5 &  \checkmark & $-$ & \cite{Lithoz_Mat}\\
    \hline
    BK7 & 2.51 & 82 & 8.3 & 1.11 & \checkmark & $-$ &\cite{BK7} \\
    Fused Silica & 2.20 & 73 & 0.52 & 1.38 &  \checkmark & $-$ &\cite{FusedSilica} \\
    ZERODUR\textregistered & 2.53 & 90 &  {$\pm$}{0.02} & 1.46 &  \checkmark & $-$ & \cite{ZERODUR} \\
    Sapphire & 3.98 & 345 & 8.4 & 27.2 &  \checkmark & $-$ & \cite{Sapphire} \\
    Kovar & 8.36 & 138 & 5.13 & 17.3 &  \checkmark & $\circ$ &\cite{Kovar} \\
\end{tabular}
\end{ruledtabular}
\begin{tablenotes}
      \item[a] Temperature stability to at least \SI{85}{\celsius} is a common requirement \cite{MILSTD883L10108A}. Application in compact high vacuum systems requires outgassing rates {\SI{< 1e-5}{\pascal\cubic\meter\per\second\per\square\meter}}. This usually involves conditioning above \SI{100}{\celsius}. In general, only a few polymers are high-vacuum compatible.
      \item[b] $++, +$ excellent/good machinability; $\circ$ demanding material, possible in standard machine shops; $-$ materials with high hardness, specialized tooling and techniques required (e.g. ultrasonic machining), expensive
      \item[c] \SI{60}{\degreeCelsius} (Rigid 4000) and \SI{101}{\degreeCelsius} (High Temp) heat deflection point at \SI{1.8}{\mega\pascal}~\cite{Rigid4000, High_Temp}.  
      \item[d] Since the value for \(\kappa\) is not available, an optimistic estimation from comparable photopolymers is stated for reference \cite{Lee2021}.
      \item[e] \SI{55}{\degreeCelsius} heat deflection point at \SI{1.8}{\mega\pascal}, \SI{61}{\degreeCelsius} glass transition temperature \cite{PLA}.
  \end{tablenotes}
\end{threeparttable}
\end{table*}

Compared to other printable materials detailed in \cref{tab:material_properties}, technical ceramics possess several unique properties: exceptionally high stiffness, low density \(\rho\), low thermal expansion, and relatively high thermal conductivity. These properties ensure mechanical stability under varying temperature conditions and mechanical loads. For applications requiring thermal insulation, materials such as \ce{ZrO2} are suitable. 
Ceramics can be manufactured into precise shapes and are compatible with a wide range of microfabrication and thick-film technologies. As electrical insulators with high temperature stability, they can be fully or partially metallized with a variety of methods. This enables realization of electrical conductor tracks, wire bonding, or soldering of components onto substrates. Combined with excellent dielectric properties, ceramics are a well-established choice for high-frequency applications. Proven processes are available for hermetically sealing ceramic materials and alumina (\ce{Al2O3}) is a standard insulator material for electrical vacuum feedthroughs. For instance, a compact vacuum chamber made of traditionally manufactured alumina with optical viewports has been successfully realized for cold atom applications \cite{Burrow2021}. In comparison to other temperature-stable materials listed in \cref{tab:material_properties}, ceramics have no magnetic susceptibility and high purity. These attributes are particularly important for magnetically sensitive applications, such as atomic magnetometers, where even minor ferromagnetic impurities in materials like aluminum alloys can disturb measurements. These advantages make ceramics an ideal choice for the development and production of robust and miniaturized quantum sensor systems, leveraging a broadly established technology base.\\
\indent One approach to realize compact optical systems and laser sources at the Ferdinand-Braun-Institut (FBH) is integrating optics, electro-optical components, and electronics on a unified substrate, a micro-optical bench (MIOB) \cite{Schiemangk2015, Kuerbis2020}. This substrate often includes custom precision reference marks for alignment, specialized shapes for component mounting and is made of ceramics (\ce{AlN}, \ce{Al2O3}) for the discussed advantages. Application of conventional methods for manufacturing ceramics presents specific challenges for micro-optical systems. The required high precision often involves additional post-processing with diamond tools, multiplying the high costs associated with ceramics. Furthermore, traditional manufacturing techniques are limited in mechanical complexity because of their inability to realize complex shapes or internal structures. They also suffer from extended lead times due to complex process chains. This becomes especially noticeable when prototyping or fabricating unique components and limits development agility. \\
\indent Additive manufacturing methods of ceramics are a versatile solution to these limitations of traditional manufacturing \cite{Chen2019}. These methods enable the creation of complex geometries and functional designs, which is advantageous in research and development (R\&D) environments. The streamlined and scalable manufacturing process enables cost-effective rapid prototyping and customization, substantially reducing lead times and enabling on-demand manufacturing. Depending on the method used, high precision and excellent surface qualities are achieved without the need for additional post-processing. Designs can be tailored to specific applications, for instance, topological optimization can be used to produce parts with reduced size and weight budgets \cite{Santoliquido2021} or to incorporate special functionality such as localized heating zones by spatially tailored thermal conduction.
Additive manufacturing presents a versatile and efficient approach for producing ceramic substrates for micro-optical systems and other components for high-performance applications in quantum technologies. Additionally, employing 3D printing enables in-house management of the entire process chain. This ensures reproducible quality and provides complete control over the manufacturing process.

\subsection*{Methods for additive manufacturing of technical ceramics}
For the generative production of ceramic parts, multiple methods with distinct benefits and specific disadvantages have been established. These include Stereolithography (SL), Binder Jetting (BJ), Fused Deposition Modeling (FDM), Direct Ink Writing (DIW) and Selective Laser Sintering (SLS). Their feedstock systems use, for example, ceramic powders or slurries, which are ceramic-filled suspensions with a photopolymer or water \cite{Chen2019}.
Lithography-based ceramic manufacturing (LCM) \cite{Schwentenwein2014} is a form of vat photopolymerization. Similar to SL, it uses light to polymerize a ceramic slurry in a vat. A dynamic mask is formed with a digital micromirror device (DMD) to shape the light and cure the polymer in defined areas. \\
\indent LCM is the method of choice for our applications, as it facilitates the manufacturing of highly complex parts with delicate features, internal structures, and high feature resolution and precision. The technique is well-suited for prototyping and small-volume productions. A selection of example components is shown in \cref{fig:example_components}.  In our experience, the quality of the final components is considerably better than printed metal or polymer materials listed in \cref{tab:material_properties}. Moreover, the mechanical and thermal properties have been demonstrated to be similar to those produced by classical manufacturing methods \cite{Schwentenwein2014, Schlacher2020, Altun2020, Rauchenecker2022}. An extensive range of materials is established, including \ce{Al2O3}, \ce{ZrO2}, \ce{AlN}, which are of particular interest for our application requirements. 
Our printer (Lithoz CeraFab S65) is equipped with a build platform measuring \(64 \times \SI{102}{\milli\meter\squared}\), provides a build height of \SI{320}{\milli\meter}, a layer thickness of \SIrange[range-units=single]{10}{100}{\micro\meter}, and a lateral resolution of \SI{40}{\micro\meter}. 
The manufacturing process is bottom-up, beginning with the dispensing of the ceramic slurry into the vat, where it is evenly distributed by a blade \cite{Schwentenwein2015}. Subsequently, the build platform is lowered into the slurry and the desired areas are selectively cured by exposure through the transparent bottom of the vat. This procedure is repeated, building the part layer-by-layer with up to \SI{150}{layers\per\hour}. Following the print process, the parts are removed, cleaned and then subjected to a thermal debinding and sintering cycle \cite{Lantada2016}. With this final step, the printed parts achieve full density. Subsequent optical quality inspection is used to optimize the design, printing and processing parameters in the next run to achieve the desired geometrical tolerances and functions. Typical process duration for the parts depicted in \cref{fig:example_components} are a few hours for printing and four to eight days for thermal post-processing, depending on the thickness of the part. Consequently, agile development cycles with periods of two weeks are possible. 

\section*{C\lowercase{er}AMR\lowercase{ef}: optical frequency reference on a printed alumina micro-optical bench}

\begin{figure}[tb]
	\centering
        \includegraphics[width=0.8\linewidth]{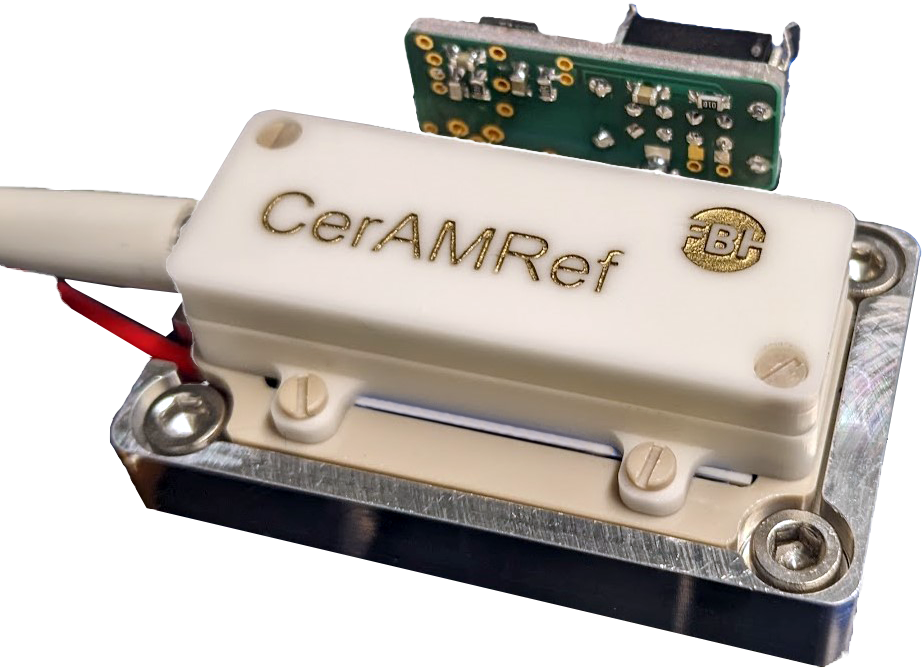}  
	      \includegraphics[width=0.85\linewidth]{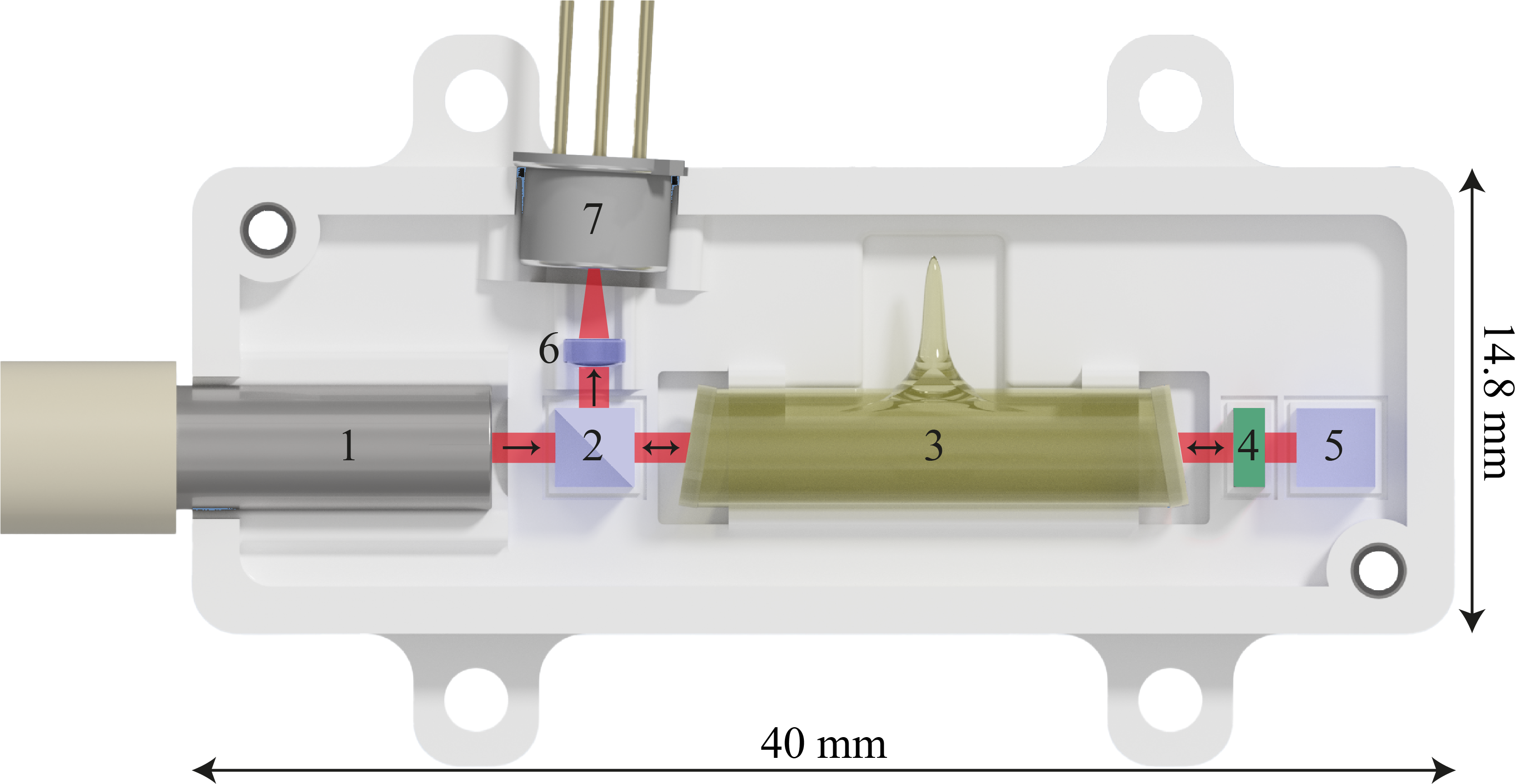}
  	\caption{\label{fig:CerAMRef+opt_layout} Top: CerAMRef frequency reference module mounted on a Peltier element with PEEK screws for thermal insulation. The PCB is electrically connected with the photodiode and Pt100 temperature sensor and includes operational amplifiers.  Bottom: a CAD rendering overlaid with the beam path of the Doppler-free FMS setup with fiber collimator (1), PBS (2), Rb-filled vapor cell (3), \( \mathrm{\lambda}/4\) plate (4), mirror (5), focusing lens (6) and photodiode (7).}
\end{figure}

To demonstrate the feasibility of additively manufactured ceramics for integrated electro-optical systems and packages, we realized a miniaturized optical frequency reference, CerAMRef, shown in \cref{fig:CerAMRef+opt_layout}. This reference employs LCM printed \ce{Al2O3} ceramics for both the micro-optical bench and the overall housing. It utilizes Doppler-free frequency-modulation spectroscopy (FMS) \cite{Bjorklund1980} on the D2 line of atomic rubidium within a vapor cell and is used to stabilize a laser system at \SI{780}{\nano\meter}. 

\subsection*{Design and assembly of the frequency reference}
\begin{figure}[t]
	\centering
	  \includegraphics[width=0.75\linewidth]{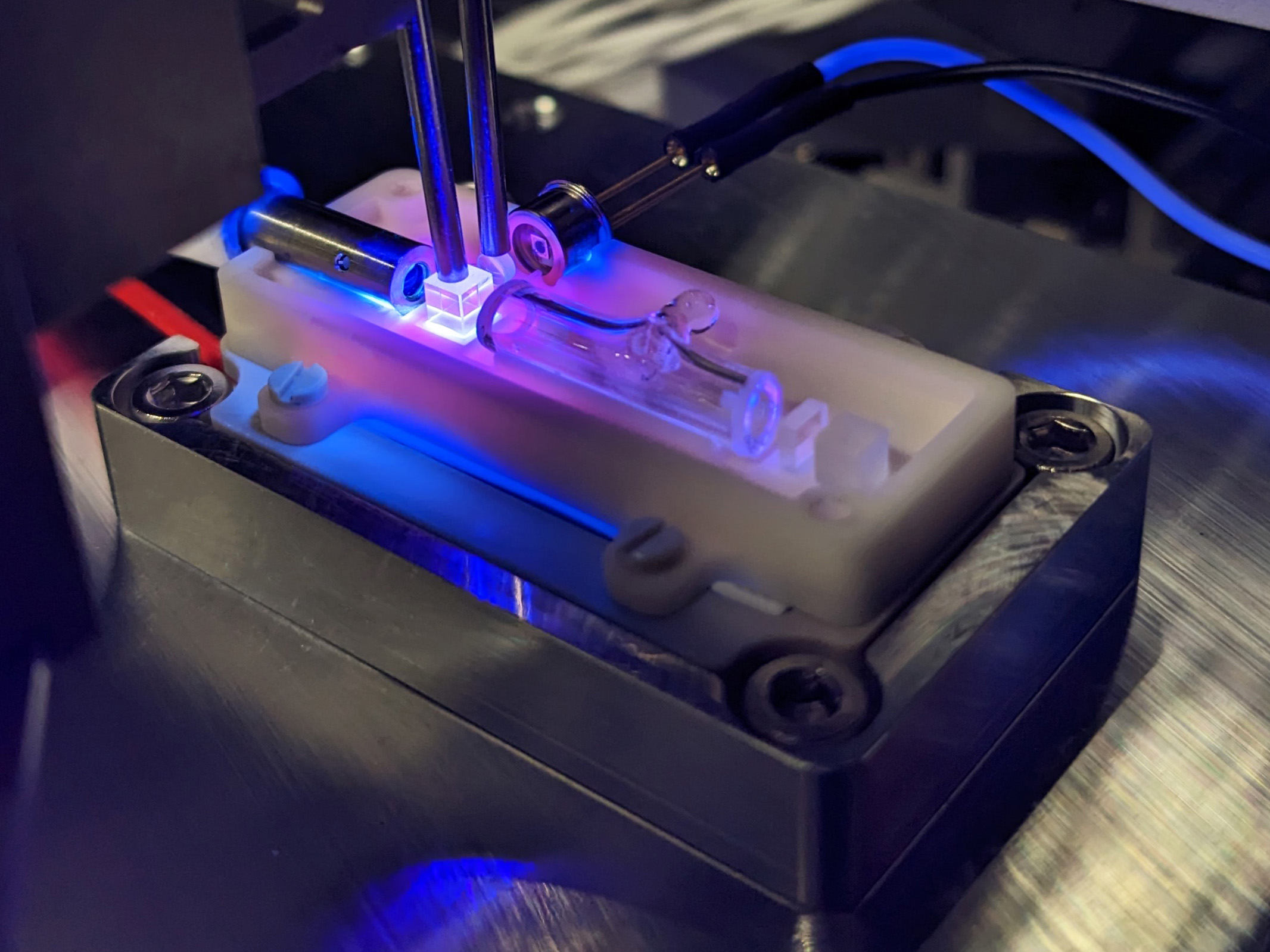}            
        \includegraphics[width=0.75\linewidth]{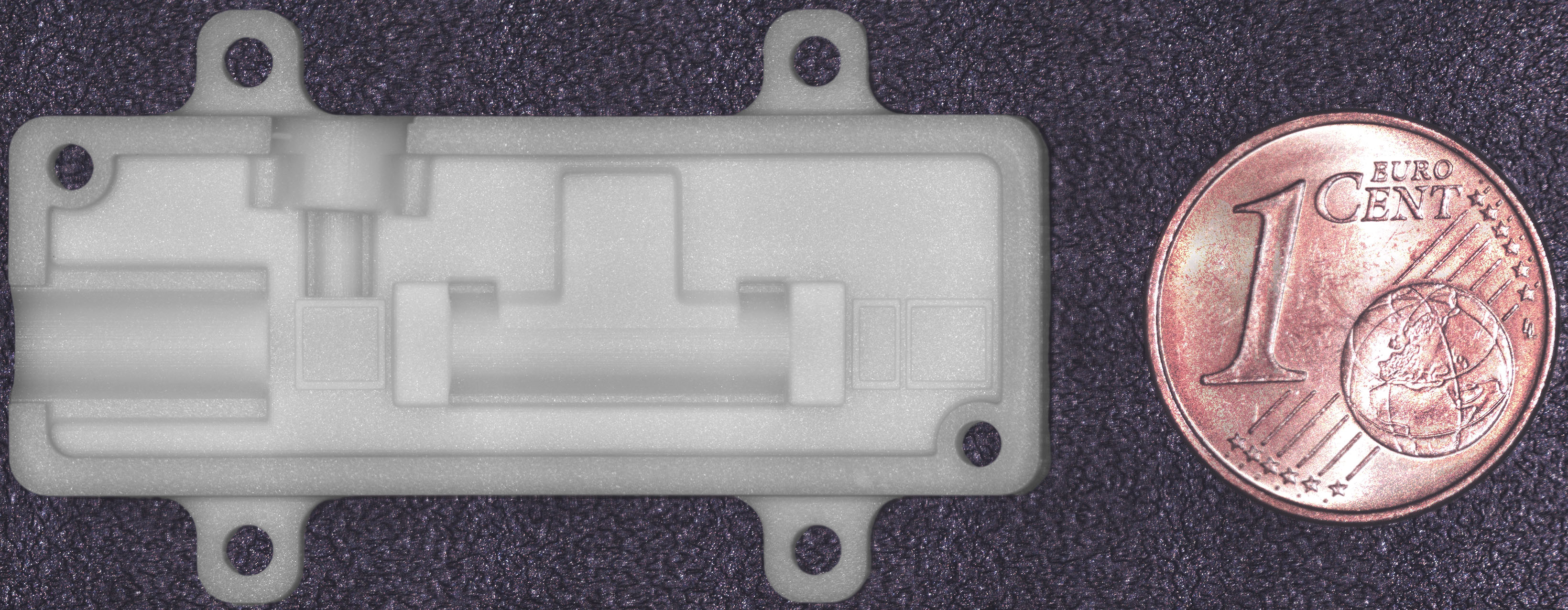}
	\caption{\label{fig:microintegration} Top: CerAMRef module during assembly in the micro-integration setup. The PBS is fixed with UV-curing adhesive after alignment while monitoring the spectroscopy signal. Bottom: microscope image of the printed alumina substrate with geometrical features for pre-aligning and mounting electro-optical components and the rubidium cell. }
\end{figure}

The optical layout is illustrated in \cref{fig:CerAMRef+opt_layout}. Light is coupled into the frequency reference with a pigtail-style fiber collimator with a single mode, polarization-maintaining fiber, emitting light with a \( 1/\mathrm{e}^2 \) beam diameter of \SI{900}{\micro\meter}. The p-polarized light first passes a polarizing beam splitter (PBS) before it enters a borosilicate cell. This cell (Rydberg Technologies) is glass-blown, \SI{15}{\milli\meter} long and \SI{3.8}{\milli\meter} in diameter, features anti-reflective (AR) coated windows angled at \SI{10}{\degree} and contains rubidium vapor in its natural abundance. The light then passes through a \(\mathrm{\lambda/4}\) plate and is reflected off a dielectric mirror aligned to overlap the incoming and reflected beam, thus enabling a Doppler-free measurement geometry. After the second pass of the \(\mathrm{\lambda/4}\) plate, the now s-polarized light is reflected by the PBS and focused onto a photodiode with a lens. This fiber-coupled design enables modular connection to the spectroscopy laser source, allowing versatile integration into various laser systems for different applications.\\
\indent The mechanical design of the frequency reference incorporates a housing and micro-optical bench made of 3D-printed alumina. This bench features custom mounting geometries, V-grooves and alignment features for the electro-optical components, which are either cylindrical, cuboidal or have an irregular shape like the glass-blown alkali cell, see \cref{fig:CerAMRef+opt_layout} and \cref{fig:microintegration}. Additionally, the housing includes a mounting recess for a Pt100 temperature sensor. The photodiode (Hamamatsu S5972) is packaged in a TO-18 housing and fixed in a custom recess on the module's sidewall. This recess is slightly tilted relative to the optical beam path to avoid etalons from spurious reflections of the diode, compare \cref{fig:CerAMRef+opt_layout}. These geometries are efficiently realized using the LCM method, resulting in a monolithic micro-optical bench that offers precise dimensions, high stiffness, and sufficient flatness after thermal processing. The processed benches achieved dimensional precision of $<\SI{50}{\micro\meter}$, flatness of \SI{30}{\micro\meter} (peak-to-valley) and surface roughness of $R_a<\SI{1.0}{\micro\meter}$ for the relevant geometries and surfaces. This is well-suited for the presented application and within the tolerances that are compensated by micro-integration. Further improvement is possible for more challenging applications by iterative optimization of the print parameters and thermal processing.
For cell heating and temperature stabilization of the overall setup, a Peltier element is placed underneath the frequency reference. A compact, custom-designed PCB, mounted on the reference module's side, is used for simple electrical connection. It connects the references photodiode and the Pt100 sensor to electrical sockets for integration into larger systems and includes a transimpedance amplifier circuit for the photodiode.
The ceramic optical frequency reference housing has dimensions of \(14.8 \times 40 \times \SI{9.5}{\milli\meter\cubed}\) and a mass of \SI{15}{\gram} including optics. The optical path length of \SI{53}{\milli\meter}, combined with the monolithic micro-optical bench and the high stiffness of alumina results in a mechanically very stable optical system without the need for realignment after integration. These specifications are a substantial improvement in size and weight compared to laboratory spectroscopy setups. Such integrated frequency references could be incorporated into higher level systems like wavemeters, optical spectrum analyzers, or cold atom sensors acting as absolute frequency reference.

The frequency reference is assembled with a specialized micro-integration facility, compare \cref{fig:microintegration}. Precision actuators adjust up to 4 components simultaneously with high precision in all 6 degrees of freedom (\SI{1}{\nano\meter} and \SI{1}{\micro\radian} stepsize). This allows for precise handling of small and delicate optics, in this application with an edge length of \SI{2.5}{\milli\meter}. During the micro-integration process, the optical system is operational, allowing for active alignment of the optical components. This is done while monitoring spectroscopy or other relevant reference signals. The components are bonded to the alumina substrate with thermal or UV-curing adhesive. Further details on the micro-integration procedures and techniques are given in \cite{Christ2023}. 

\subsection*{Performance demonstration}
\begin{figure*}[t]
    \centering
    \includegraphics[width=0.8\linewidth]{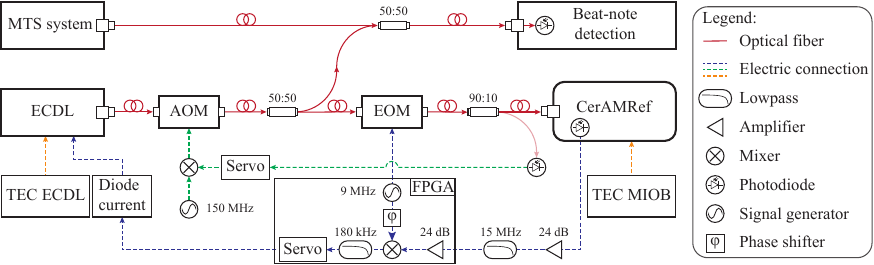}
    \caption{\label{fig:perf_eval}Experimental setup for spectroscopy, laser frequency stabilization and frequency instability measurements of the CerAMRef reference with optical connections (red) and electrical connections for frequency (blue), laser intensity (green) and temperature (orange) stabilization. Further details on the electronic components are provided in the text.}
\end{figure*}

\begin{figure}[tb]
    \centering
    \includegraphics[width=\linewidth]{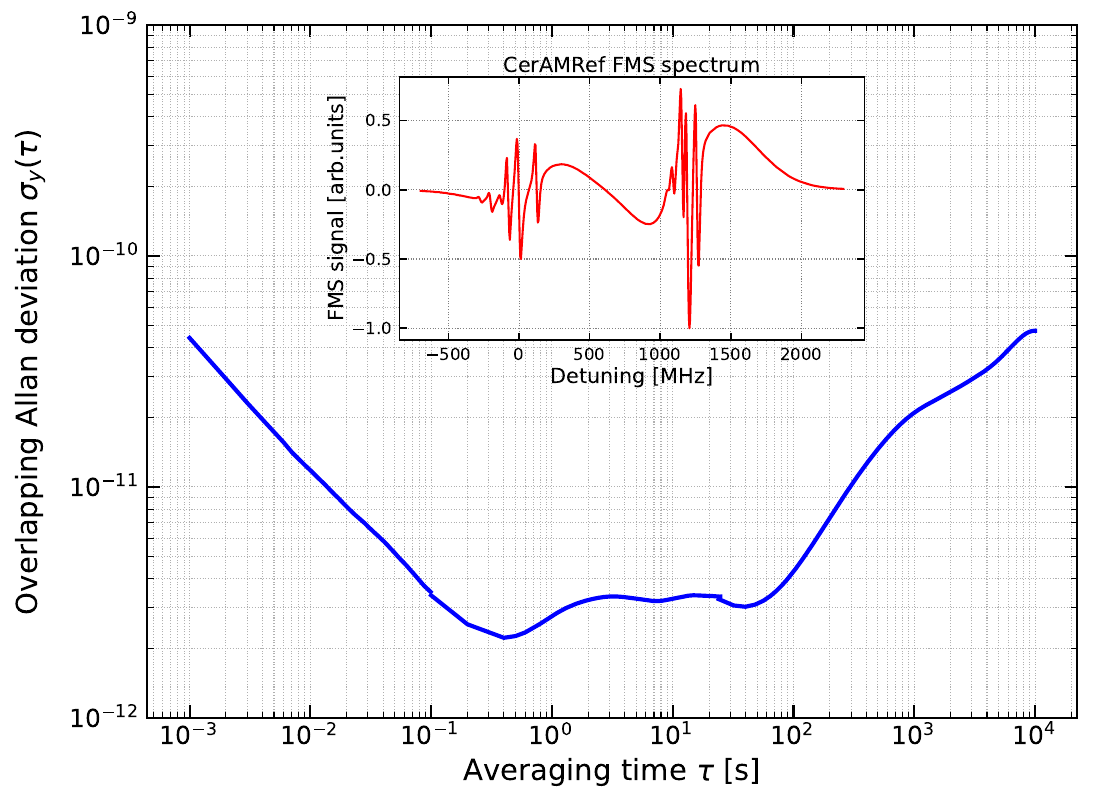}
    \caption{\label{fig:adev}Overlapping Allan deviation of the CerAMRef, calculated from three beat-note time series. The insert displays the FMS spectrum of the CerAMRef using the locked-to transition as reference for the detuning. This spectrum exhibits features from the \ce{^87Rb} D2 \(F=2 \rightarrow F'\) and \ce{^85Rb} D2 \(F=3 \rightarrow F'\) transitions.}
\end{figure}

To assess the performance of the CerAMRef module, we set up a laser system to perform frequency modulation spectroscopy and determine the frequency instability.
The experimental setup is illustrated in \cref{fig:perf_eval}. The CerAMRef is operated at \SI{780}{\nano\meter} with a micro-integrated, narrow linewidth external-cavity diode laser (ECDL) developed at FBH \cite{Kuerbis2020, Doeringshoff2019}. To stabilize its temperature, Peltier elements and a thermoelectric cooler controller (TEC, Meerstetter Engineering TEC-1091) are used. The temperature of the CerAMRef module is also controlled by a Peltier element and stabilized at \SI{30}{\degreeCelsius}.
The light from the fiber-coupled ECDL laser passes a fiber-coupled acousto-optical modulator (AOM, G\&H Fibre-Q SFO5498), operated at \SI{150}{\mega\hertz} and used for stabilization of the optical power, and is then split with a 50:50 fiber splitter. One part is used for beat-note measurements \cite{Saleh1991} with a reference system,  while the other is coupled to the spectroscopy setup for frequency stabilization. This part interfaces with a fiber-coupled electro-optical phase-modulator (EOM, iXblue NIR-MPX800-LN) that modulates the laser light at a frequency of \SI{9}{\mega\hertz}. Another fiber splitter after the EOM splits off 10\,\% of the light, which is guided to a photodiode (Thorlabs PDA36A2). The photodiode signal is used for power stabilization through a servo loop~\cite{Wiegand2022}, operated on a field-programmable gate array (FPGA, Red Pitaya STEMlab 125-14), which is connected to the AOM. The remaining 90\,\% of the light is connected to the CerAMRef module, compare \cref{fig:CerAMRef+opt_layout}. The signal of the reference's photodiode is subsequently amplified, low-pass filtered and relayed to an FPGA based signal processing system (Liquid Instruments Moku:Pro). Here, the signal is shifted in phase and demodulated with the modulation signal of the EOM, which is also generated by the FPGA. This demodulated signal is used as error signal and internally processed by a servo to generate a control signal, providing feedback to the ECDL's current modulation port and stabilize the lasers frequency to the target transition. The FMS spectrum of CerAMRef is shown in \cref{fig:adev}, exhibiting \ce{^85Rb} and \ce{^87Rb} D2 line transitions.

The frequency instability of the laser system locked to the CerAMRef is determined from a beat-note with a reference laser system of  a transportable quantum gravimeter \cite{Schmidt2011}. It employs modulation-transfer spectroscopy (MTS) \cite{Shirley82} of the \ce{^85Rb} D2 \(F=3 \rightarrow F'=4\) transition. The CerAMRef's system is frequency-locked to the \ce{^87Rb} D2 \(F=2 \rightarrow F'=2/3\) crossover transition, which shows the best slope and signal to noise ratio for the lock. For this setting we measure the expected beat-note frequency of \SI{1220}{\mega\hertz} including a \SI{40}{\mega\hertz} offset from an AOM in the MTS reference system \cite{Rb85, Rb87}.

To evaluate the frequency instability of the CerAMRef, the beat-note signal between the locked laser systems is recorded with a frequency counter (Pendulum CNT-91) and analyzed using the overlapping Allan deviation \(\sigma_y(\tau)\) \cite{Allan1966, Rutman1991}. The Allan deviation provides a measure of the relative frequency instability, defined as  \(y=\Delta f/f_0\) with frequency fluctuations \(\Delta f\) and optical frequency \(f_0\) (here \SI{384}{\tera\hertz}), over various averaging times \(\tau\). 
\cref{fig:adev} shows the Allan deviation of the beat-note frequency between the two lasers locked to the CerAMRef and the MTS reference, calculated from three dead-time free time series. A linear drift of \SI{6.16}{\hertz\per\second} is subtracted from the data. For averaging times below \SI{1}{\second} we find white frequency noise at a level of \( 1 \times 10^{-12} / \sqrt{\tau} \). At \SI{1}{\second} averaging time, an overlapping Allan deviation of \num{3e-12} is achieved, increasing to \num{4e-12} at \SI{100}{\second}. The presented stability is typical for a rubidium D2 FMS reference and comparable to previous studies \cite{Gruet2017, Zi2017}. For averaging times between \SI{10}{\milli\second} and \SI{300}{\second} the Allan deviation shows an instability well below \num{1e-11} (\SI{3.8}{\kilo\hertz}) and for averaging times up to \SI{1e4}{\second}, well below \num{1e-10} (\SI{38}{\kilo\hertz}), perfectly suitable for cold atom applications using magneto-optical traps (MOT) \cite{Sabulsky2020, Li2023}. 

\section*{Discussion and outlook}

This work underlines the significant potential of additively manufactured ceramics as key technology for miniaturized functional designs of electro-optical systems and compact quantum technologies. Ceramics are characterized by a high Young's modulus, low density, high temperature stability, and thermal conductivity that ranges from isolating (as for \ce{ZrO2}) to very high (as for \ce{AlN}), compare \cref{tab:material_properties}. They close a technology gap in available materials for additive manufacturing and enable the fabrication of compact, lightweight components and assemblies. As a demonstrator, we have realized a micro-integrated optical frequency reference based on FMS spectroscopy of the rubidium D2 line at \SI{780}{\nano\meter}. This demonstrator achieves significant improvement in size and weight budget compared to lab-based setups through optical micro-integration techniques on an additively manufactured alumina substrate. Its performance is well-suited for mobile, atom-based quantum sensing applications. With a volume of \SI{6}{\milli\liter} and mass of \SI{15}{\gram} for the integrated spectroscopy module, the CerAMRef is the smallest fiber-coupled FMS reference to our knowledge. Further reductions can be achieved by incorporating a MEMS-based vapor cell \cite{Liew2004a}, a more compact collimator or by integrating electronic circuits and components on the electrically insulating optical bench. The frequency reference can be easily adapted for other wavelengths by switching the atomic species in the vapor cell (e.g. potassium, caesium) and thus address a broad application range. Furthermore, the optical layout of the reference could be extended to integrate a second collimator and optical path, realizing an MTS setup. This could further enhance the stability at longer averaging times, while leading to only a small increase of the system volume.

\begin{figure}[tb]
	\centering
    \includegraphics[width=0.8\linewidth]{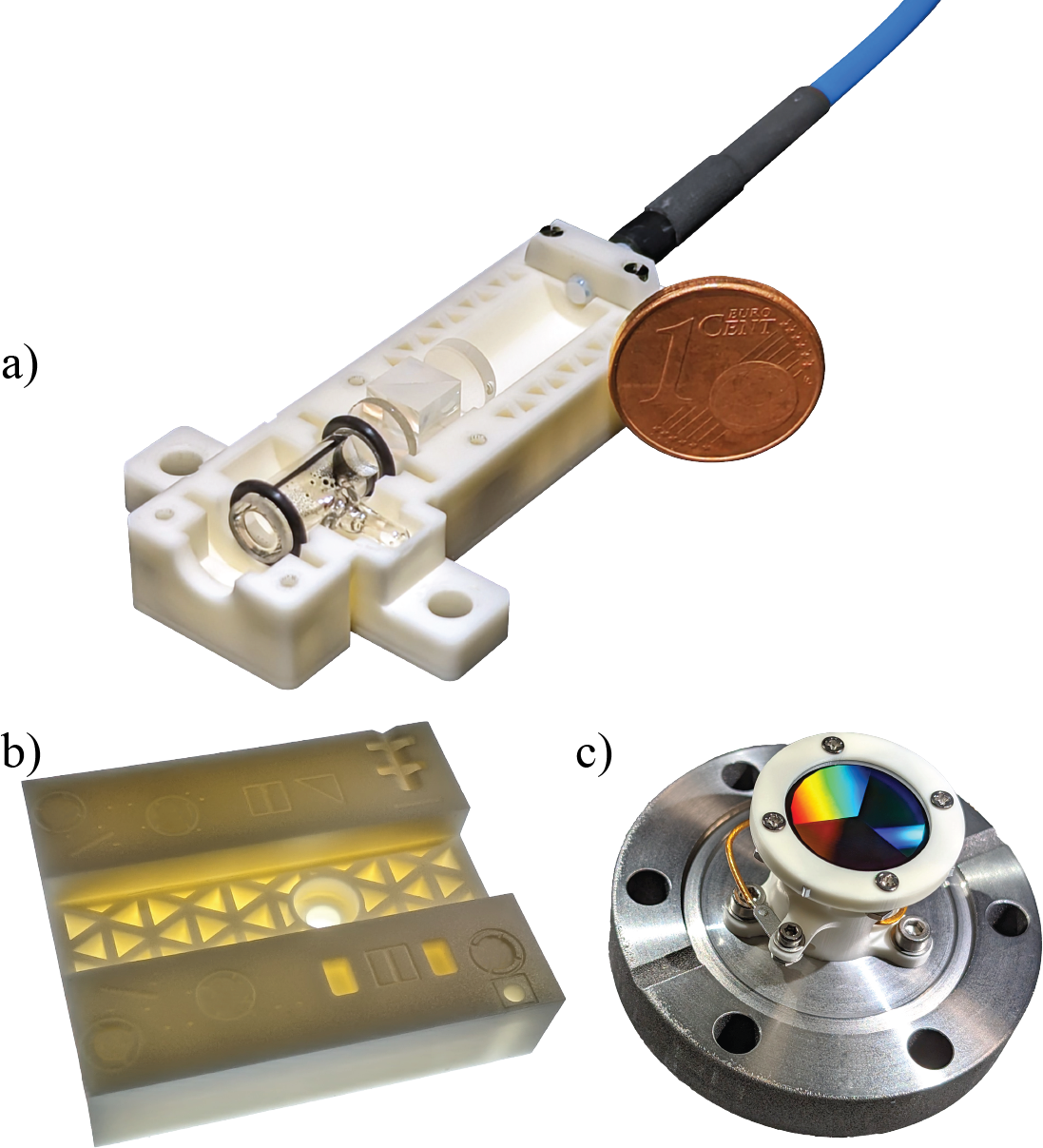}
	\caption{\label{fig:future_applications} Future applications of additive manufactured alumina in quantum technologies: a) depicts an atomic magnetometer featuring a heated caesium vapor cell and micro-integrated optics. In b), a micro-optical bench is shown (\qtyproduct[product-units = power]{36 x 43 x 10}{\milli\metre}), designed for a miniaturized crossed beam optical dipole trap. The transmitted light highlights the lightweight design with internal lattice structure. c) displays an ultra-high vacuum application: an alumina holder with a three-fold surface grating for a grating MOT, mounted on a printed aluminum CF40 flange. The electrically isolating and thermally stable ceramic enables easy integration of restively heated dispensers and planar magnetic structures.}
\end{figure}

Further applications of the technology currently investigated at FBH include compact atom-based magnetometers and atom trapping layouts.
Atomic magnetometers use optical excitation of vaporized alkali metal atoms to create spin alignment which is sensitive to external magnetic fields. This enables precise magnetic field detection \cite{Budker2013}. \cref{fig:future_applications} shows a miniaturized physics package of an optically pumped magnetometer (OPM), micro-integrated on an additively manufactured ceramic bench. The sensitivity of an OPM strongly depends on the temperature of the alkali vapor \cite{Tiporlini2013} and hence relies on suitable thermal stability and localized heating. Regarding these demands, additive manufacturing of ceramics is promising to extend the range of field applications for atomic magnetometers.

An optical bench designed for a micro-integrated crossed-beam optical dipole trap \cite{Grimm1999, Christ2023} is depicted in \cref{fig:future_applications}. Such a trapping scheme can be used to cool and manipulate neutral atoms for ultra-cold atom applications \cite{Perrin2009}. The lightweight design of the optical bench with several mounting geometries for optical elements and internal lattice structures benefits from the high stiffness of alumina. Additionally, \cref{fig:future_applications} depicts a printed ceramic holder for a three-fold diffraction surface grating, which is used to realize a grating magneto-optical trap (gMOT) \cite{Cotter2016}. This holder is attached to a printed aluminum CF40 flange with an electrical feedthrough. In this application, the vacuum compatibility, thermal stability and electrical insulation properties of \ce{Al2O3} are essential. It simplifies the electrical connections for heating a rubidium dispenser and accommodates planar electrical structures beneath the grating chip for magnetic field generation \cite{Chen2022_MOT}. Further applications in quantum technologies can be addressed by different ceramic materials, such as \ce{AlN} for high-power diode laser systems requiring advanced thermal management.

R\&D applications of additively manufactured ceramics using vat photo-polymerization are continually evolving and future innovations and improvements of the printing technology and materials are to be expected. One current development is multi-material printing \cite{Schlacher2021}. This technique facilitates combinations of different ceramics (e.g. \ce{Al2O3} and \ce{ZrO2}) and also ceramics and metals in a single print. These combinations can be various: layer-by-layer, within individual layers, or as functionally graded materials. This approach allows for the spatial tailoring of mechanical and thermal properties to meet specific application requirements. A particularly promising aspect of this technology is the integration of electric conductors. This could facilitate the creation of components with embedded capabilities for magnetic field generation or heating elements, leading to a further system integration and reducing the number of components and production steps.

\section*{Acknowledgements}
The authors thank S. Schwertfeger for support in printer operation, J. Kluge for support in frequency measurements, A. Hahn for microscope images, A. Ukhanova for contribution of the grating holder, M. Schoch, the prototype engineering lab and F. Ertl (Lithoz GmbH) for technical support.

This work is supported by the German Space Agency (DLR) with funds provided by the Federal Ministry for Economic Affairs and Climate Action (BMWK) due to an enactment of the German Bundestag under grant number 50WM2070 (CAPTAIN-QT), 50WM1949 (KACTUS-II) and 50WM2169 (MYOQUANT).

\section*{Author contributions}
M.C. envisioned the application of printed ceramics, designed and assembled the CerAMRef, set up, conducted and evaluated the frequency measurements, and drafted the manuscript; M.C. and C.Z. designed and set-up the micro-integration facility; C.Z. supported the assembly of the CerAMRef; S.N. operated the ceramic printer together with M.C., processed the parts, designed the magnetometer, and assembled it with C.Z.; B.L. operated the MTS reference and supported the frequency measurements; K.D. advised on instrumentation, measurement layout, and performance optimization of the frequency measurements; M.C. and M.K. acquired funding and conceptualized the work. All authors reviewed the manuscript. 

\section*{Conflict of interest}
The authors declare no conflict of interest.

\section*{Data availability statement}
The data that support the findings of this study are available from the corresponding author upon reasonable request.

\section*{Keywords}
additive manufacturing, technical ceramics, miniaturized optical systems, miniaturized physics packages, quantum technology, optical frequency reference, vacuum system integration

\bibliography{references}

\end{document}